\documentclass[aps,prl,twocolumn,groupedaddress]{revtex4}

\usepackage{graphicx}
\newcommand{\be}{\begin{equation}}
\newcommand{\ee}{\end{equation}}

\begin{document}


\title{Bose-Einstein Condensation of Molecular Hydrogen in Nanotube Bundles}
\author{Francesco Ancilotto$^{1,2}$, M. Mercedes Calbi$^1$, Silvina M. Gatica$^1$ and Milton W. Cole$^1$}
\affiliation{$^1$Physics Department, Pennsylvania State University, University Park, Pennsylvania 16802; \\ $^2$ INFM (UdR Padova and DEMOCRITOS National Simulation Center, Trieste, Italy) \\ and Dipartimento di Fisica ``G. Galilei'',
Universit\`a di Padova,\\ via Marzolo 8, I-35131 Padova, Italy.}

\date{\today }

\begin{abstract}

We evaluate the effects of heterogeneity on the density of states of H$_2$ molecules 
inside interstitial channels within bundles of carbon nanotubes. As temperature 
(T) falls, the density increases within those tubes having the greatest binding 
energy. At T $\approx$ 10 mK, the molecules undergo Bose-Einstein condensation, 
exhibiting a singular heat capacity.

\end{abstract}
\pacs{}
\maketitle

The subject of adsorption within bundles of carbon nanotubes has received 
considerable attention recently, owing to both its fundamental interest and potential 
applications (e.g. gas storage and isotope separation). One focus of the research is 
one-dimensional (1D) and quasi-1D phases of matter. These include condensing and 
crystallizing phases of buckyballs within tubes, He and H$_2$ within the interstitial 
channels (IC's) between tubes and various gases within grooves on the outside of the 
nanotube bundles \cite{qsiev,bob,cro,oscCv,rmp,aldo,boro,c60,bienf,carlo,hecond}. Most 
analyses (including those of our group) have assumed 
that the tubes are identical and parallel, forming an ordered lattice. Real nanotube 
bundles, in contrast, consist of a disordered array of tubes with a distribution of 
radii. A logical question arises: how reliable are predictions that ignore such a 
variable environment? Shi and Johnson have recently shown that predictions 
incorporating heterogeneity agree better with adsorption data than do the 
idealized models \cite{hete}. Stimulated by their work, we have explored the behavior of 
quantum fluids (He and H$_2$) in such a nonuniform environment at low temperature (T). 
In this paper, we describe an intriguing result: Bose-Einstein condensation (BEC) 
of H$_2$ molecules occurs as {\em a consequence of the heterogeneity}. This paper makes 
predictions about this phenomenon that are testable experimentally. Similar behavior 
is expected for $^4$He atoms. We note that a superfluid phase of para-H$_2$ in confined 
geometries has been proposed, but experimental evidence of it is lacking thus far 
\cite{h2bec}.

The occurrence and properties of the BEC transition are determined by the density 
of states $N(E)$ of the constituent particles. Thus, an initial focus is the determination 
of that function, derived from the single particle energy spectrum. We assume that 
particles do not interact (based on the assumption of low density and weak inter-IC 
interactions). A typical IC presents a highly confining geometry for the molecules 
if the bundle is close-packed. Because of this confinement (in the x-y plane, 
perpendicular to the tubes), the molecules have a large zero-point energy of motion. 
Theoretical values \cite{interc} of this energy are of the order of 500 K. This result is 
consistent with an enormous isotopic heat difference at low coverage (which equals the
binding energy difference) between D$_2$ and H$_2$ in a nanotube bundle observed 
experimentally \cite{oscar}. That difference was found to be about 
250 K, a factor $\approx$ four larger than the difference found on the graphite 
surface \cite{vid} (due to less localization in that case).  

Our analysis assumes that molecules can move within the IC's in order to achieve 
chemical equilibrium with a coexisting vapor phase. At low T, because of the 
confinement, only the lowest energy state of transverse motion is excited. We 
call this state's energy $E_t({\bf R})$, where ${\bf R}=(R_1,R_2,R_3)$ is a 
vector whose components are the radii of the tubes surrounding a particular IC. 
We have evaluated $E_t({\bf R})$, with interesting results. The calculation 
assumes that motion parallel to the IC is that of free molecules, so the total 
energy of a particle with (z component of) momentum p is given by $E(p,{\bf R})=
E_t({\bf R}) + p^2/(2m)$. If the potential is corrugated, one must replace the 
particle's mass $m$ with its band mass \cite{hecond,band}. The density of states 
for H$_2$ is obtained by summing over the IC's present in the given sample and 
integrating over p:

\begin{eqnarray}
N(E)&=&\sum_{{\bf R},p}\; \delta[E - E(p,{\bf R})] \nonumber \\		
    &=& \frac{L}{\hbar \pi} \, \left(\frac{m}{2}\right)^{1/2} \int_0^E \; 
dE_t \,\frac{g(E_t)}{\sqrt{E - E_t}}  	
\end{eqnarray}

Here $g(E_t)$  is the density of states for the transverse oscillation problem 
and $L$ is the length of the tubes. If the length is not constant, the variation 
can be included in $g(E_t)$.

The energy $E_t({\bf R})$ is evaluated from the potential energy $V({\bf r,R})$ 
of the molecule at position {\bf r}. To compute $V$, we add contributions 
from the three neighboring tubes, ignoring corrections from more distant tubes 
and many-body effects associated with the screening of van der Waals interactions 
by the adjacent tubes \cite{milen}. Although these approximations introduce some errors 
in the values of $E_t$ they do not affect the key predictions of this work, which 
are sensitive to the variation of $E_t({\bf R})$. The potential from each neighboring 
tube was derived with the method of Stan et al \cite{uptake}. Because $V({\bf r,R})$ 
varies 
rapidly with {\bf r}, values of $E_t({\bf R})$ include large anharmonic and 
small anisotropic contributions.

The form of $N(E)$ for a given collection of nanotube bundles depends on sample 
preparation. A sample is represented by an ensemble of points (one for each IC) 
in {\bf R} space. The density of points in {\bf R} space, a function $f({\bf R})$, 
is defined so that $f({\bf R})\,d{\bf R}$ is the number of IC's within an 
infinitesimal volume $d{\bf R} = dR_1\,dR_2\,dR_3$, centered at {\bf R}. The function
 $f$ enters the transverse density of states through this expression:

\be
g(E) = \int \,d{\bf R}\, f({\bf R})\, \delta[E - E_t({\bf R})]
\ee

At very low T, we need the value of $f({\bf R})$ only in the immediate vicinity 
of the global minimum of the energy ($E_m$ at ${\bf R}_m$), but at higher T 
the specific experimental distribution affects the results quantitatively. 
Here,  we have assumed that the IC's are uniformly distributed in {\bf R} 
space within a radius spread of width 3 \AA$\,$ near ${\bf R}_m$. More general 
results will be described in a complete report of this work . We find, as one 
might expect, that ${\bf R}_m$ occurs along the diagonal, symmetry line 
($R_1=R_2=R_3$). Along this line, which we call the (1,1,1) line, there occurs 
a global minimum $E_m=-1052.97$ K at {$R_i$}=9.95 \AA. To derive $g(E)$, 
it is important to know 
the variation of $E_t({\bf R})$ near this minimum. Consider a particular 
displacement (in the $R_2-R_3$ plane) from ${\bf R}_m$ to a neighboring point 
for which $R_1$ has the same value, while $R_2$ and $R_3$ are slightly different: 
$R_2 = R_1+ \delta$ and $R_3=R_1 - \delta$. This change, parallel to the (0,1,-1) 
direction, yields an extremely small increase in the H$_2$ energy, indicative 
of a very slow variation of the function $E_t({\bf R})$ near ${\bf R}_m$. This 
behavior, shown in Fig.1 reveals a long valley of low energy states in this 
direction; the same behavior occurs along the five equivalent directions, e.g. (1,0,-1).

\begin{figure}
\includegraphics[height=3in]{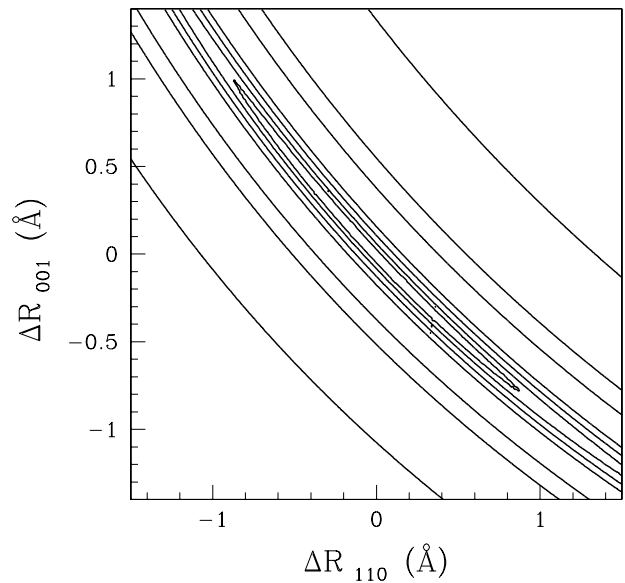}
\caption{Contour plot showing the variation $E_t({\bf R})-E_m$ near the 
minimum ${\bf R}_m$ (which is chosen as the origin). From the contour closest to the 
minimum (the closed one) to the more distant ones, the energy contours correspond to 
$E_t({\bf R})-E_m = 0.005, 0.01, 0.05, 0.1, 0.5, 1$ and 5 K.}
\end{figure}

The transverse density of states $g(E)$, from Eq. 2, is shown in Fig. 2. Note 
that $g(E)$ is proportional to $\sqrt{E - E_m}$ for small $E - E_m$; the 
prefactor is determined by the principal axes of curvature of the function 
$E_t({\bf R})$. This square root behavior is identical to that found near a van 
Hove singularity in the phonon density of states of a 3D system near a minimum 
in the Brillouin zone, for the same reason- phase space topology \cite{ziman}. 
At higher energy, instead, the behavior of $g(E)$ switches to $1/\sqrt{E - E_m}$. 
This (-1/2) power law follows from the fact that the higher energy displacements 
from ${\bf R}_m$ are quasi-1D. As seen in Fig.1, the iso-energy contours are perpendicular 
to the diagonal, so that the energy gradient in {\bf R} space is along the diagonal, 
with essentially constant transverse variation, a 1D situation.

\begin{figure}
\includegraphics[height=4in]{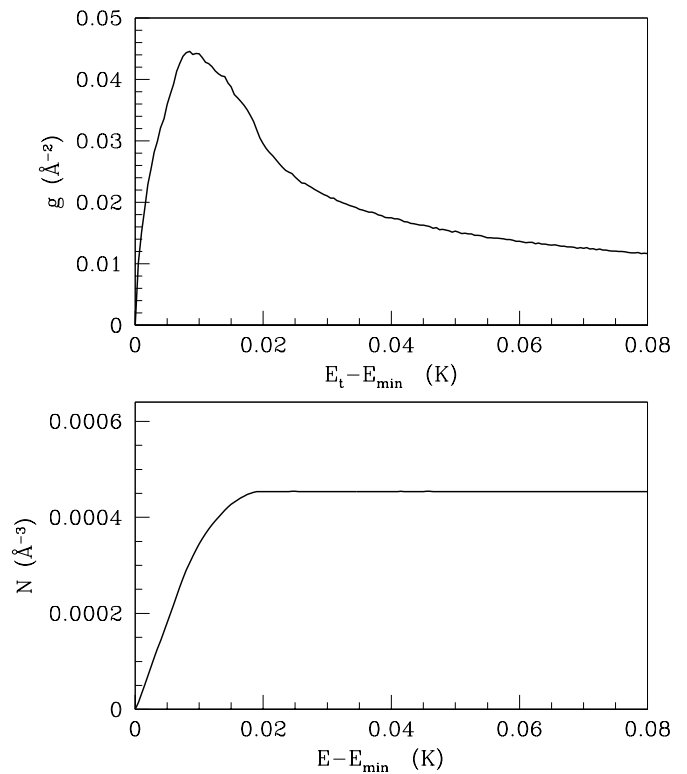}
\caption{Upper panel: Calculated transverse density of states $g(E)$.
Lower panel: Total density of states $N(E)$, obtained from Eq.1 in the text.
$E_m=-1052.97$ K is the lowest energy level corresponding to $R_1=R_2=R_3=9.95$ \AA.}
\end{figure}

Fig. 2 also presents $N(E)$, derived with Eq. 1, which convolutes the transverse 
spectrum with the 1D motion along the axis. The resulting power law behavior 
can be understood from realizing that if $g(E)$ is proportional to $(E-E_m)^n$, 
for some $n$, then $N(E)$ is proportional to $(E - E_m)^{n+1/2}$ . Hence, we find 
that $N(E)$ is proportional to $(E-E_m)$ near threshold. This linear behavior is 
that characteristic of a 4D gas in free 
space; the result implies that {\em this system exhibits 4D gas behavior at low T}. For 
$E>20$ mK above threshold, instead, $N(E)$ becomes approximately constant, 
corresponding to the density of states  of a 2D gas. Thus, there arises a 
dimensionality crossover originating from the anomalous transverse density 
of states. We emphasize that the 4D regime is a direct consequence of the 
existence of a minimum in the function $E_t({\bf R})$, a result that is not sensitive 
to the details of the calculation \cite{groove}.  

The thermodynamic behavior of the system is derived with the usual bose gas theory. 
For a given number of molecules, $N$, the chemical potential $\mu$ is determined 
from the relation 

\be
N = \int \,dE \,\frac{N(E)}{e^{\beta(E-\mu)} - 1}
\ee

At a specified $T=1/(k_B \beta)$, this relation yields a maximum value $N_{max}$ 
when $\mu$ equals the lowest energy of the system, $E_m$, at which point BEC 
begins. That is, a macroscopic fraction of the molecules fall into the lowest 
energy state when $N > N_{max}$. Equivalently, at fixed $N$, BEC occurs when T 
falls below the inverse function $T_c = T(N_{max})$. The resulting dependence 
on $N$ of $T_c$ is shown in Fig. 3. As seen there, $T_c$ is of order 10 mK, which 
is experimentally accessible. Fig. 4 shows the specific heat $C_N(T)$, calculated 
from the energy as a function of $\mu$ and T. The novel behavior observed in 
the figure is a result of the unusual form of $N(E)$. At low T, $C_N$ is 
proportional to $T^2$, a consequence of the 4D (linear) variation of $N(E)$ at 
low $E$. Note the presence of singular behavior of $C_N(T)$ as one approaches 
the transition from above and a cusp at $T_c$ itself. At relatively high T, 
$C_N(T)/(N k_B)$ is essentially unity because $N(E)$ is 2D-like at high $E$; 
a nondegenerate 2D gas has $C_N(T)/(N k_B)=1$. Behavior for $T>0.2$ K, 
not shown, is very sensitive to the distribution of nanotube radii.

\begin{figure}
\includegraphics[height=2.5in]{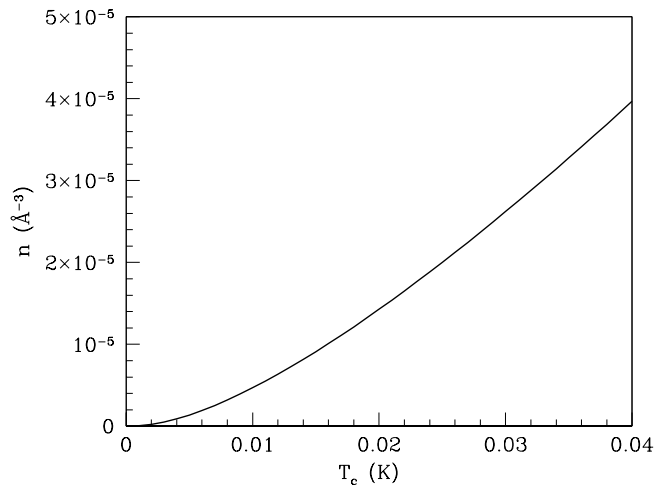}
\caption{Density of H$_2$ molecules as a function of BEC transition temperature.}
\end{figure}

\begin{figure}
\includegraphics[height=2.5in]{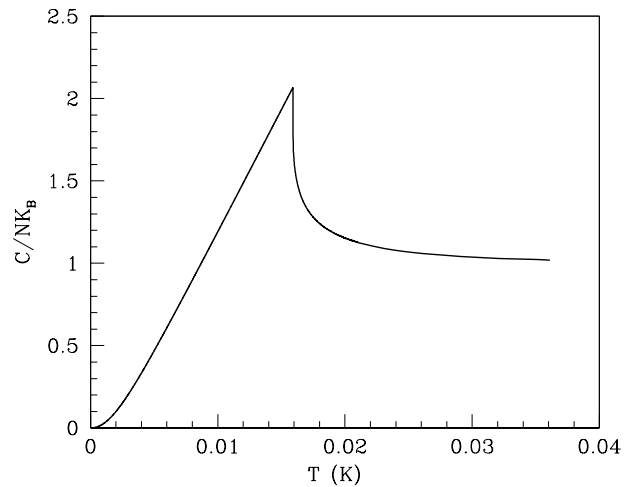}
\caption{Temperature dependence of the specific heat (per molecule) 
for molecular density $n=1\times 10^{-5}$ \AA$^{-3}$.}
\end{figure}

Discussions of these results with colleagues have led to several questions 
concerning the transition. One is this: since each IC contributes a density 
of states characteristic of a 1D system, why does non-1D behavior arise here? 
The answer is that particles can exchange between different IC's because 
of their common access to the vapor region. In practice, this may be a very 
slow process, leading to nonequilibrium behavior. The presence of breaks or 
holes in the tubes may alleviate this kinetic problem without invalidating 
the model, since heterogeneity is an essential aspect of the model. Another 
question is whether similar behavior occurs for other bose gases. Indeed, 
BEC of $^4$He is predicted by a similar analysis to occur if the sample's 
distribution of tubes includes those with some near its energy minima in {\bf R}
 space, which occur near 8.5 \AA. Finally, one might wonder about the 
effects of interparticle interactions, which have been ignored up to this point 
\cite{he4}. Indeed, some previous studies of H$_2$ in IC's have found that 
a liquid-vapor 
condensation occurs at a higher temperature ($\approx$ K) (in the absence 
of heterogeneity) \cite{hecond}. A very recent study, however, found that 
nanotubes' screening of the intermolecular interaction reduces $T_c$ to about 
10 mK \cite{milen}. However, that calculation omitted the role of heterogeneity, 
which is relevant, 
according to the Harris criterion, since the 3D specific heat critical exponent 
is positive \cite{harris}. We expect that this condensation temperature is further 
reduced by disorder, enabling the BEC transition to occur. 

We summarize our results as follows. Heterogeneity alters the qualitative 
behavior of the low energy spectrum of H$_2$ molecules.  The lowest-lying states 
of the system are those of particles in that channel. As T falls, particles 
aggregate 
in the (essentially 4D) space of quantum states, ({\bf R},p), with bose 
statistics having a dramatic effect, i.e. BEC, below a transition temperature 
of order 20 mK. Anomalous behavior is predicted for the specific heat, a 
consequence of the unusual density of states, which is 4D-like at very low 
energy and 2D-like at somewhat higher energy. An experimental probe of the 
real-space molecular density should reveal the needle-like concentration, 
below $T_c$, of a macroscopic fraction of the particles within the lowest 
energy channel. 

Most intriguing to us is that this transition is a direct consequence of 
disorder, since the perfectly uniform system of identical nanotubes yields 
strictly 1D, nonsingular behavior. Such a dramatic effect of heterogeneity 
occurs elsewhere in low temperature physics. Examples include the spin-glass 
transition \cite{schiffer}, the effect of tunneling states on thermal behavior of 
glasses \cite{glasses} and the effects of disorder on monolayer films \cite{hedisorder}.

This research has benefited from discussions with Moses Chan, Vin Crespi, 
Susana Hern\'andez, Jainendra Jain, Peter Schiffer, Paul Sokol, Flavio Toigo 
and David Weiss and support from NSF and the Hydrogen Storage program at Penn State.
F.A. acknowledges funding from MIUR-COFIN 2001.


\begin{references}
\bibitem{he4} Recall that the 3D ideal gas BEC transition of $^4$He is preempted 
by the interaction-driven condensation to a liquid below 5.2 K, resulting in 
a ground state condensate fraction about 10\%.
\end{references}
\end{document}